# Investigation on organic magnetoconductance based on polaron-bipolaron transition

W. Qin, S. Yin, K. Gao and S.J. Xie[*]

*School of Physics, State Key Laboratory of Crystal Materials, Shandong University, Jinan 250100, People's Republic of China*

**Abstract**

We explore the magnetoresistance (MC) effect in an organic semiconductor device based on the magnetic field related bipolaron formation. By establishing a group of dynamic equations, we present the transition among spin-parallel, spin-antiparallel polaron pairs and bipolarons. The transition rates are adjusted by the external magnetic field as well as the hyperfine interaction of the hydrogen nuclei. The hyperfine interaction is addressed and treated in the frame work of quantum mechanics. By supposing the different mobility of polarons from that of bipolarons, we obtain the MC in an organic semiconductor device. The theoretical calculation is well consistent to the experimental data. It is predicated that a maximum MC appears at a suitable branching ratio of bipolarons. Our investigation reveals the important role of hyperfine interaction in organic magnetic effect.

---

[*] Corresponding author e-mail address: xsj@sdu.edu.cn

**I. Introduction**

Recently, there has been growing interest in organic magnetic field effect (OMFE) in non-magnetic organic semiconductor devices. This interest is motivated by organic materials' intrinsic magnetic field-related physical properties [1]. Organic magnetoconductance (OMC) effect, as one aspect of OMFE, refers to the phenomenon in which conductance changes as a function of external magnetic field ($B$) and is observed in relatively low magnetic field (lower than 100 mT) at room temperature [2-6]. Since the OMC effect is highest known among all non-magnetic materials, this effect can be readily used for organic magnetic sensors and magnetically controlled organic optoelectronic devices. Understanding the mechanism of OMFE may not only improve the general understanding of field response in organic materials but also help in the development of future application.

To understand the physical mechanism of OMFE, a large number of bipolar [7-10] and unipolar [11-15] organic devices have been fabricated and investigated up to now. A unipolar device means that there is no luminescence [11, 13], but an apparent OMC signal is found. For example, in unipolar polymer device ITO/PFO/Au [16], where only holes are injected from anode ITO, a magnetoresistance (reverse MC) about -5% was reported and value of magnetoresistance decreased with increasing applied voltages. In unipolar device PEDOT/PFO/Au [14], a MC of 1% was obtained. And in unipolar organic small molecular device ITO/Alq$_3$/Au [13], a value of MC could be as high as 10%. It has been summarized that the OMC traces have a certain universality: the dependence of measured OMC value on external magnetic field $B$ can be well fitted with either the fully saturated Lorentzian function $\text{MC}(B) \propto B^2/(B^2+B_0^2)$ or the weak saturated non-Lorentzian function $\text{MC}(B) \propto \left[B/(|B|+B_0)\right]^2$ upon the concrete organic materials or devices. As the injected carriers may exist in the form of both polarons and bipolarons, by supposing that polaron hopping rate between sites are spin related or the external magnetic field and the hyperfine interaction related, Bobbert *et al.* [17, 18] proposed steady-state rate equation for polarons and spinless bipolarons. With a Gaussian distribution of the hyperfine effective field, through the Monte Carlo simulation, they obtained the OMC and showed that OMC is caused by blocking of the current by certain sites at which bipolaron formation can take place. In this paper, we will consider the quantum effect of the hyperfine interaction. We try to establish a group of dynamic equations for spin-parallel polarons, spin-antiparallel polarons and bipolarons. These equations include the transition rate between polarons and bipolarons, which is closely related with both the external magnetic field and the hyperfine interactions of the hydrogen nuclei. By considering different mobility or velocity of a polaron from that of a bipolaron, we calculate the OMC in a unipolar organic device and compare the results with possible experimental data. This paper is organized as follows. In Sec. II, we will describe our theoretical analysis, establishing the model and deriving the main calculation formulas. The main results and discussions are given in Sec. III. In addition, a comparison between our calculation and experimental result is provided in Sec. III. Finally, we present our conclusions in Sec. IV.

## II. MODEL AND METHOD

In an organic device, the injected electrons (or holes) are trapped by the molecular deformation, forming localized polarons or bipolarons. A polaron holds one electron (or hole) with spin ±1/2 (in units of $\hbar$), while a bipolaron holds two electrons (or holes) with no spin. They transport in an organic device and act as carriers. The total current density is given by $j = en_p v_p + 2en_{bp} v_{bp}$, where $n_p(n_{bp})$ is the polaron (bipolaron) density, $v_p(v_{bp})$ is the polaron (bipolaron) velocity, and $e$ is the elementary charge. We consider that a bipolaron moves slower than a polaron or $v_{bp} < v_p$ [19].

Within the organic layer, when a polaron moves near to another polaron, they will interact and have a probability to form a bipolaron. The large on-site exchange effects will lead to that the energy of a spin triplet bipolaron is much higher than that of a singlet one. Hence, we assume that the bipolaron can only appear as a spin singlet. In order to give a clear description on the transition between polarons and bipolarons, we divide the injected carriers as spin-parallel polaron pairs, spin-antiparallel polaron pairs and singlet bipolarons. Among them, transition takes place between spin-parallel polaron pairs and spin-antiparallel polarons pairs due to spin mixing, which is related to the magnetic field and hyperfine interactions, and transition between spin-antiparallel polaron pairs and bipolarons due to the electron-lattice confinement interaction. A schematic diagram of the transitions is shown in Fig. 1.

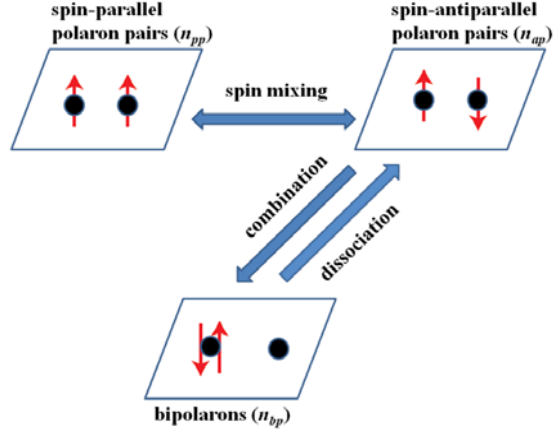

FIG. 1. (Color online) A schematic diagram of transition for spin-parallel polaron pairs, spin-antiparallel polaron pairs and bipolarons

Introducing the transition rate $\gamma_{ap}$ to describe the hopping from a spin-parallel polaron pair to a spin-antiparallel one and $\gamma'_{ap}$ for its reverse process, we write the dynamic equations of these carriers,

$$\frac{dn_{pp}}{dt} = -\gamma_{ap} n_{pp} + \gamma'_{ap} n_{ap} \tag{1a}$$

$$\frac{dn_{ap}}{dt} = +\gamma_{ap} n_{pp} - \gamma'_{ap} n_{ap} + k n_{bp} - b n_{ap} \tag{1b}$$

$$\frac{dn_{bp}}{dt} = -k n_{bp} + b n_{ap}, \tag{1c}$$

where $n_{pp}(n_{ap})$ is the spin-parallel (-antiparallel) polaron pair density and $n_p = 2(n_{pp} + n_{ap})$, and $n_{bp}$ is the density of bipolaron. $-\gamma_{ap} n_{pp}$ ($+\gamma'_{ap} n_{ap}$) means the decreasing (increasing) of the spin-parallel pair density due to the spin-mixing. The parameter $b$ describes the local recombination rate for two polarons of opposite spin forming a bipolaron, while $k$ describes the reverse process, where a bipolaron decomposes into two polarons of opposite spin [20, 21].

To get the transition rate $\gamma_{ap}$ ($\gamma'_{ap}$), we have to consider the spin related interactions in the organic layer. In an organic semiconductor, a polaron (or bipolaron) is tightly confined at one or a few of molecules due to the strong electron-lattice interactions. In this case, the interaction between the localized polaron spin $\hat{\vec{S}}$ and the hydrogen nuclei spins $\hat{\vec{I}}$ will become apparent. When an external magnetic field is applied, the total Hamiltonian for a pair of polarons is written as,

$$\hat{H} = \sum_{i=1}^{2} g\mu_B B \hat{S}_{z,i} + a \hat{\vec{I}}_i \cdot \hat{\vec{S}}_i, \tag{2}$$

where $\mu_B$ is the Bohr magneton, $g$ the Lande factor, and $a$ denotes the strength of the hyperfine interaction. For simplicity we do not consider the spin correlation between the polaron spins. Except at the eigenstate, state $|s_P, I_H\rangle_I |s_P, I_H\rangle_{II}$ will evolve with time. For example, for the pair with spin-parallel state $|1\rangle = |\uparrow, \Uparrow\rangle_I |\uparrow, \Downarrow\rangle_{II}$ at beginning, it has been derived that the total spin will revolute and its z-component is given by $S_z(t) = \frac{1}{2} + \frac{1}{2}(p_{pp}^1 - p_{ap}^1)$ [14], where $p_{pp}^1 = \frac{1}{2}\left(1 + \frac{\omega^2}{\omega^2+a^2} + \frac{a^2}{\omega^2+a^2}\cos\sqrt{\omega^2+a^2}\,t\right)$ and $p_{ap}^1 = 1 - p_{pp}^1$ mean the probabilities for finding the pair in a spin-parallel and spin-antiparallel state, respectively, and $\omega = g\mu_B B$ is a parameter which is related to external magnetic field. Their time averages are given by $\bar{p}_{pp}^1 = \frac{1}{2}\left(1 + \frac{\omega^2}{\omega^2+a^2}\right)$ and $\bar{p}_{ap}^1 = \frac{a^2}{2(\omega^2+a^2)}$. Considering all possible nuclear spin conformations in state $|s_P, I_H\rangle_I |s_P, I_H\rangle_{II}$, we obtain the probability of spin-parallel state $\bar{p}_{pp} = \frac{1}{16}\sum_{j=1}^{16}\bar{p}_{pp}^j = \frac{1}{2} + \frac{\omega^4}{16(\omega^2+a^2)^2}$ and that of spin-antiparallel one $\bar{p}_{ap} = 1 - \bar{p}_{pp}$.

Now let us consider $N_p$ pairs of spin states. Due to the external magnetic field effect and hyperfine interaction, transition will be taken place between the spin-parallel pairs and spin-antiparallel pairs. At the equilibrium state, we have $\gamma_{ap} n_{pp} = \gamma'_{ap} n_{ap}$, where $n_{pp} = \bar{p}_{pp} N_p$ and $n_{ap} = \bar{p}_{ap} N_p$, $N_p = n_{pp} + n_{ap}$ is the total polaron pair density. From these equations, we obtain the transition rate from spin-parallel pairs to spin-antiparallel ones $\gamma_{ap} = \frac{1}{2}[1 - \frac{\omega^4}{8(\omega^2+a^2)^2}]\gamma_0$, and the rate from spin-antiparallel pairs to spin-parallel ones $\gamma'_{ap} = \frac{1}{2}[1 + \frac{\omega^4}{8(\omega^2+a^2)^2}]\gamma_0$, where $\gamma_0$ is a parameter. If there is no the external orientated magnetic field, we have $\gamma_{ap}(B=0) = \gamma'_{ap}(B=0) = \gamma_0/2$.

With the transition rate, by solving Eq. (1) we can get the redistribution of the carrier density and its evolution when external magnetic field is applied. Then the MC is calculated from $j(B) = 2e[n_{pp}(B) + n_{ap}(B)]v_p + 2en_{bp}(B)v_{bp}$,

$$\text{MC} = \frac{j(B) - j(0)}{j(0)} \times 100\%. \tag{3}$$

## III. RESULTS AND DISCUSSIONS

Let us consider an organic unipolar device. The injected carriers exist as polarons and bipolarons.

When the magnetic field is absent, They arrive at an equilibrium distribution, which are determined by Eq. (1): $n_{pp}^0 = n_{ap}^0 = N(2+b/k)^{-1}$ and $n_{bp}^0 = N(1+2k/b)^{-1}$, $N = n_{pp} + n_{ap} + n_{bp}$. If carrier density is small in organic semiconductors, there has no much chance that two polarons meet each other, so the majority is single charged polaron. But under a high bias, with the increasing of injected carrier numbers, the added carriers in organic layer bearing single charged polarons lead to the direct transition of polaron to bipolaron state(only for spin-antiparallel pairs) [22]. And in this case, the majority may become bipolaron. By setting the parameter $k$ and $b$ or $k/b$, we may choose the majority is polaron or bipolaron. When external magnetic field is applied, again by solving the Eq. (1), we obtain the evolutions of the carrier redistributions with applied magnetic field $B$=100 mT, which are showed in Fig. 2. The hyperfine interaction parameter $a$ is given to correspond to an effective field $B_{hf} = a/g\mu_B = 3.5$ mT [23]. Figure 2(a) is the case that bipolarons are the majority while Fig. 2(b) the minority.

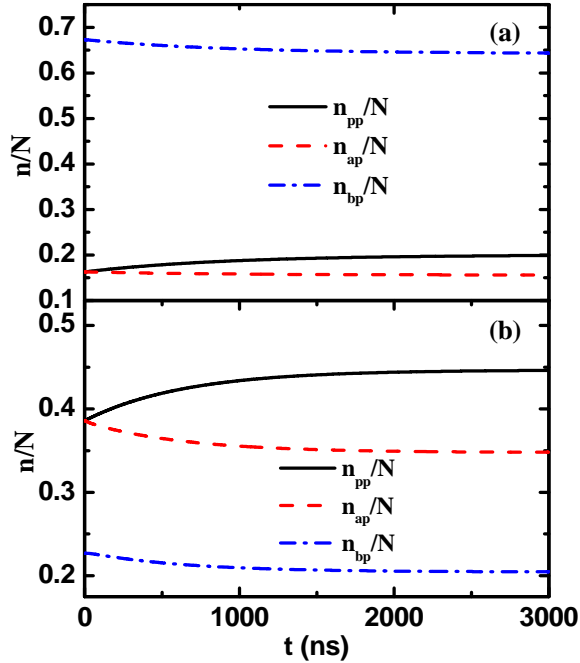

FIG. 2. (Color online) The evolution of bipolarons, spin-antiparallel polaron pairs and spin-parallel polaron pairs, external magnetic field is 100 mT. The parameters are set as: $\gamma_0 = 2\times10^6 s^{-1}$, (a) $k$=1.7×10$^7$ s$^{-1}$, $b$=7×10$^8$ s$^{-1}$ and (b) $k$=1.7×10$^7$ s$^{-1}$, $b$=1×10$^7$ s$^{-1}$.

It is found that polarons and bipolarons will redistribute when external magnetic field is applied. No matter whether the bipolaron is majority or minority, the densities of bipolarons and spin-antiparallel polarons will decrease. At the same time, the density of spin-parallel polarons will increase. It shows that the magnetic field will increase the transition from bipolarons to polarons. A new equilibrium distribution will be reached after a duration of 2000 ns when the magnetic field is switched on in present parameters.

When the new equilibrium distribution of polarons and bipolarons is reached under external magnetic field and the hyperfine interaction, let us calculate the current density $j = en_p v_p + 2en_{bp} v_{bp}$ through the organic device and then the MC. For a weak magnetic field as in the actual experimental cases, we have,

$$MC = MC(n, B) + MC(v, B), \quad (4)$$

where $MC(n,B) = \frac{\Delta n_p v_p + \Delta n_{bp} v_{bp}}{n_p v_p + n_{bp} v_{bp}}$ and $MC(v,B) = \frac{n_p \Delta v_p + n_{bp} \Delta v_{bp}}{n_p v_p + n_{bp} v_{bp}}$ are the magnetoconductances separately due to the change of carrier density and velocity. We consider that carrier density variation is mainly responsible to the OMC effect. For example, by using of the electroluminescence spectroscopy and charge-induced absorption spectroscopy technique, Nguyen *et al.* measured the dependence of polaron densities on the applied magnetic field and found that the density increases with the magnetic field [24, 25]. Therefore, here we only consider $MC(n, B)$. By solving Eq. (1) and setting $t \to \infty$, we obtain the MC of an organic device,

$$MC = MC_\infty \frac{\omega^4}{\omega^4 + 2\beta a^2 \omega^2 + \beta a^4}, \quad (5)$$

where $MC_\infty = \frac{2(1-\alpha)k/b}{(\alpha + 2k/b)(7 + 16k/b)}$ is the saturated MC value. $\alpha = v_{bp}/v_p$ is the velocity ratio between bipolaron and polaron. $\beta = 1 + \frac{1}{16k/b + 7}$. From Eq. (5) it is found that the MC behavior is tightly related to the magnetic field and the hyperfine interaction. As $\beta \approx 1$, when the external magnetic field is stronger than the hyperfine effective field, $B \gg B_{hf}$ (or $\omega \gg a$), we have $MC \approx MC_\infty \frac{\omega^2}{\omega^2 + 2\beta a^2} = MC_\infty \frac{B^2}{B^2 + 2\beta B_{hf}^2}$, which is a fully saturated trace and is well fitted to the Lorentzian form to explain experimental data. However, if the external field is small, $B < B_{hf}$ or the values of $B$ and $B_{hf}$ are close to each other, it is found that the MC will deviate from the Lorentzian function. The whole evolution of MC with the magnetic field is shown in Fig. 3. The inset in Fig. 3 gives the MC behavior under a small magnetic field and the Lorentzian function $MC \propto B^2/(B^2 + B_0^2)$ as a comparison. From Eq. (5) we also obtain that the saturated MC is mainly determined by the velocity ratio $\alpha$. In organic semiconductors, it has been indicated that a polaron and a bipolaron have different reorganization energy, which means that a polaron will have different velocity or mobility from a bipolaron. Usually, a bipolaron moves slower than a polaron, so a positive MC will appear in a unipolar device. Of course, if a bipolaron moves faster than a polaron ($\alpha > 1$), a negative MC is obtained. Taking suitable parameters we obtain the experimental value about 2% of the saturated MC [13]. Especially, it is found that the saturated MC value is independent of the hyperfine interaction, although the hyperfine interaction is extremely vital for the appearance of OMC effect. This conclusion seems to be consistent with the experimental observation of OMFE in organic devices fabricated separately with protonated and deuterated DOO-PPV (for example, see Fig. 3c in Ref. 23).

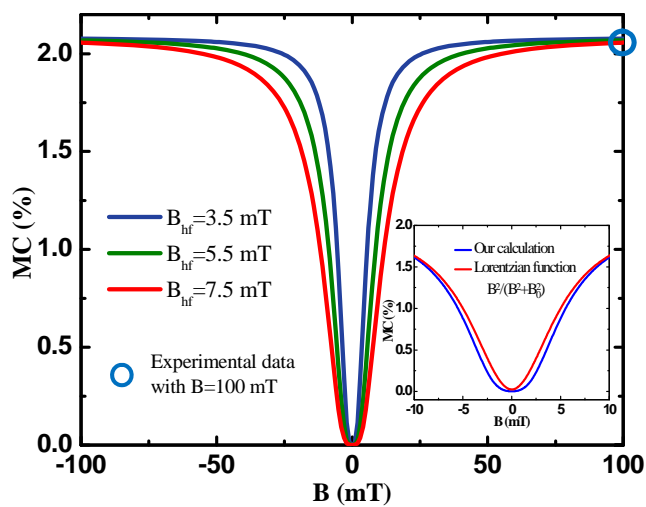

FIG. 3. (Color online) MC curves with different hyperfine interaction, i.e. $B_{hf}$=3.5, 5.5 and 7.5 mT separately. Symbol ○ is the experimental data at external magnetic field $B$=100 mT [13]. The inset shows the comparison of MC between our calculation and Lorentzian function $MC \propto B^2 / (B^2 + B_0^2)$ under a small external magnetic field, $B_{hf}$=3.5 mT, $B_0$=5.1 mT. The other parameters are the same as those in Fig. 2 (a).

We also find that the branching ratio or the portion of bipolarons has an important effect on OMC. The portions of polarons and bipolarons in the organic layer are affected by two factors: one is the external applied bias. As mentioned above, under a low bias, only a small quantity of carriers are injected, in this case there has no much chance that two polarons meet each other, so the majority may be single charged polaron. But with increasing the applied voltage, additional charges are injected into organic semiconductor and the polaron configuration will be transformed to bipolaron configuration by the addition of one extra charge to it [22]. So the majority is bipolaron. The other factor is the concrete organic material. In small molecules, such as pentacene and Alq$_3$, injected electrons (or holes) form polarons localized on molecular sites. Bipolaron is not easy to form due to the strong Coulomb repulsive interaction when two polarons are confined at only one molecular site [26, 27]. But in polymers, the situation is different. Two polarons may be confined together to form a singlet bipolarons in a long polymer chain. It has been widely studied that, even including the Hubbard electron-electron interaction, a bipolaron is still energetically more favorable to form [28, 29]. Therefore, we may suppose that, in small molecule materials, bipolarons are the minority, while in polymers, bipolarons are majority (especially under a high bias). Here we study the dependence of MC on the branching ratio of bipolarons, and the result is showed in Fig. 4. The inset in Fig. 4 gives the experimental data of unipolar organic small molecular device PEDOT/Alq$_3$/Au [30] and polymer device ITO/PFO/Au [16] under different applied bias.

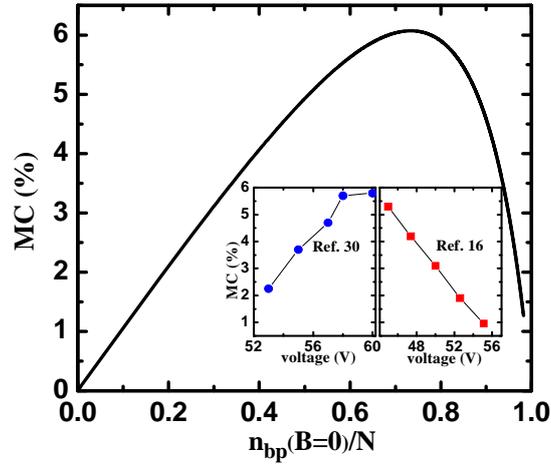

FIG. 4. (Color online) External magnetic field $B$=100 mT and $\alpha$ =0.15, MC as a function of bipolaron ratio.

It is showed that, in the case of small branching rate of bipolarons, the MC increases with the rate, while in the case of large branching rate, the MC decreases with the rate. It has been reported that, in small molecular device PEDOT/Alq$_3$/Au [30], the MC increases with the applied voltage or injected carrier density (as shown in the inset of Fig. 4, left). Combining the above analysis, we speculate that bipolarons are the minority in Alq$_3$. But in polymer device ITO/PFO/Au [16], it was reported that the MC decreases with the applied voltage (as shown in the inset of Fig. 4, right). As polarons extend to be confined into bipolarons in polymers, we speculate that bipolarons are the majority in PFO.

## IV. CONCLUSION

In conclusion, a group of dynamic equations based on the transition between bipolarons and polarons are proposed to explain MC effect in organic devices. We treat the hyperfine interaction within quantum mechanism rather than a classical effective field approximation. By considering all possible configurations of the hydrogen nuclei spin states, we give the transition rate between a spin-parallel polaron pair and spin-antiparallel one. It is found that the transition process is irreversible when the external field is included. And this transition rate is closely related to the hyperfine interaction. Then we investigate the redistribution of polarons and bipolarons when the magnetic field is applied and calculate the MC in an organic device. It is found that the results are well consistent to experimental data. Especially, the line shape of the MC($B$) is closely related with the hyperfine interaction, which shows the importance of hyperfine interaction in OMC. It is also obtained that the branching rate of bipolarons plays a vital role for the OMC behavior. With some experimental reports, we analyze the relationship between the MC and the portion of bipolarons in different organic devices.

**ACKNOWLEDGEMENTS**

The authors would like to acknowledge the financial support from the National Basic Research Program of China (Grant No.2010CB923402 and No.2009CB929204) and the National Natural Science Foundation of the People's Republic of China (Grant No.11174181 and No.21161160445).